\documentclass{aa}  

\usepackage{graphicx}
\usepackage{txfonts}
\usepackage{booktabs}
\usepackage{pdflscape}
\usepackage{longtable}  
\usepackage{placeins}
\usepackage{float}
\usepackage[colorlinks=false, linkcolor=blue]{hyperref}
\usepackage{afterpage} 
\usepackage{orcidlink}
\usepackage{lipsum}

\begin{document}

\title{An updated census of the Vast Polar Structure member galaxies}
\titlerunning{VPOS member galaxies}
\authorrunning{A. M. Martínez-García}
\author{Alberto Manuel Martínez-García\inst{\ref{IAA}}\orcidlink{0000-0001-9755-3872}\corrauth{ammartinez@iaa.csic.es}}
\institute{Instituto de Astrofísica de Andalucía, CSIC, Glorieta de la Astronomía, 18008 Granada, Spain\label{IAA}}
\date{Received XXXXXX; accepted XXXXX}
 
\abstract{
The Vast Polar Structure (VPOS) is a thin, planar arrangement of dwarf galaxies oriented  almost perpendicular to the Milky Way (MW) disc. Since its discovery, the census of VPOS member galaxies has progressively expanded  thanks to continuous findings of new dwarfs in the vicinity of the MW and increasingly precise measurements  of their phase-space coordinates. In this work, we present the largest census to date of VPOS member galaxies, building on a sample of 73 MW dwarfs  for which positional and kinematic data are available in the literature. Based on their orbital poles, we identify 22 VPOS members.
We note that the majority of VPOS member galaxies are co-rotating, while only three systems are counter-rotating (Bootes III, Kim 1, and Sculptor). Our revised census increases the known VPOS member population by nearly 50\%, introducing   Delve 4,  Kim 1, Kim 3, Eridanus III, Pictor II, and YMCA-1 as new on-plane dwarfs. From this updated census, we revise the structural parameters of the VPOS, confirming its planar ($c/a =0.27$) and remarkably thin  (15 kpc) nature. 
}


\maketitle
\nolinenumbers

\section{Introduction}
Fifty years ago, it was reported that several of the Milky Way (MW) dwarf satellites known at the time, such as the Magellanic Clouds, Draco, Sculptor, and Ursa Minor, some globular clusters, and the Magellanic Stream, lay along a great circle around our Galaxy (\citealt{Kunkel1976, Lynden-Bell1976}). Since then, the progressive discovery of dwarf galaxies in the vicinity of the MW, and the analysis of their distribution, confirmed the presence of a thin planar arrangement of multiple MW satellites nearly perpendicular to the Galactic disc. This arrangement is commonly known as the Vast Polar Structure (VPOS; e.g. \citealt{Pawlowski2012, Pawlowski2013}).
Through pioneering measurements of the proper motions (PMs) of MW satellites, and particularly with the advent of the \textit{Gaia} mission (\citealt{GaiaCollaboration2016}), it has also been confirmed that most of the VPOS galaxies are co-rotating (\citealt{Metz2007, Metz2008, PawlowskiKroupa2013, PawlowskiKroupa2018,Fritz2018, Li2021, Taibi2024, MartinezGarcia2025}), that is, they share the same sense of motion within the plane. Beyond the MW, analogous planes of satellites have been detected around several galaxies, such as M31 (\citealt{Conn2013, Ibata2013}), M81 (\citealt{Chiboucas2013}), M101 (\citealt{Merritt2014}), or NGC 4490 (\citealt{Karachentsev2024}). However, it remains challenging to assess the kinematic coherence of these planes given their distance to the MW and the current lack of reliable PMs.

The cosmological relevance, and fit of planes of satellites in the $\Lambda$ cold dark matter ($\Lambda$CDM) paradigm has been widely debated. Initial claims of differences between $\Lambda$CDM expectations for satellite distributions and the observed anisotropy of the VPOS (\citealt{Kroupa2005}) sparked the so-called plane of satellite galaxies problem (\citealt{Bullock2017, Pawlowski2018}), which is itself debated (\citealt{Sawala2023}).
Searches for analogous satellite distributions in cosmological simulations have greatly struggled to find planes of satellites, reporting them to be rare and transient 
(e.g. \citealt{Lovell2011, Ibata2014, Pawlowski2014, Pawlowski2018, PawlowskiKroupa2018, SantosSantos2020, Samuel2021, Pawlowski2021, Pawlowski2024}). further reinforcing these tensions. However, several recent works claimed to have more frequently found persistent, kinematically coherent planes of satellites in higher-resolution simulations (zoom-in and cosmological) using novel planarity measurement methods (\citealt{SantosSantos2023, GamezMarin2025, Madhani2026}), suggesting  that such structures can indeed arise within the  $\Lambda$CDM framework.

The origin of the VPOS and of planes of satellites in general is equally puzzling. While multiple formation scenarios have been proposed to date, no definitive solution has yet been reached (see \citealt{Pawlowski2018}). These formation  mechanisms include the anisotropic accretion of satellites along filaments of the large-scale structure (\citealt{Libeskind2005, Zentner2005, Lovell2011, Libeskind2011, Libeskind2014, Libeskind2015, Wang2020b}), the accretion of groups of satellites (\citealt{LyndenBell1995, DOnghia2008, LiHelmi2008, Smith2016}, but see \citealt{Metz2009b}), the flattening of the local cosmic web (\citealt{GamezMarin2024}), and the formation of tidal dwarf galaxies from the debris of interactions between gas-rich galaxies (\citealt{Pawlowski2011, Pawlowski2012, Hammer2013, Akib2025}). More recently, it was also reported that massive Gaia–Sausage–Enceladus-like mergers (\citealt{Helmi2018b, Belokurov2018}) in MW analogues might give rise to coherent and persistent planes of satellites (\citealt{RodriguezCardoso2026}). 

Beyond $\Lambda$CDM, the formation of planes of satellites has also been explored in alternative cosmological frameworks (e.g.  modified Newtonian dynamics), where simulations have suggested that the planes of the MW and M31 might have formed via tidal dwarf galaxies generated as a result of an early MW--M31 flyby (e.g. \citealt{Bilek2018, Bilek2021, Banik2018, Banik2022},  see also \citealt{Collins2015, Taibi2024}). This event is only possible in a dark-matter-free context. 

The Large Magellanic Cloud (LMC) has also been proposed as a main cause in the formation of the VPOS. On the one hand, it might have induced the clustering of orbital poles of the MW satellites during its recent passage (\citealt{GaravitoCamargo2021}, but see also \citealt{CorreaMagnus2022, Pawlowski2022}). On the other hand,  if the LMC were in its second pericentre about the MW, the VPOS might have formed from LMC satellites that were stripped during its earliest passage (\citealt{Vasiliev2024}).

Ultimately, our understanding of the cosmological significance, formation, and evolution of planes of satellites depends on our ability to accurately constrain the properties of the nearest and most accessible example: the VPOS. Over the past few decades, the census of MW satellites has expanded substantially (\citealt{Doliva-Dolinsky2025}), while the availability and quality of their kinematic data have dramatically improved through the \textit{Gaia} Early Data Release 3 (EDR3; \citealt{GaiaEDR3}). The unprecedented wealth of positional and kinematic data currently available for dwarf galaxies in the vicinity of the MW provides an ideal foundation to revisit the membership and structure of the VPOS.

We present the largest census to date of VPOS member galaxies based on a sample of 73 MW dwarfs, whose phase-space coordinates are available in the literature. From this updated census, we present revised structural parameters for the VPOS.

\section{Data and methods}
\label{sec:methods}
\subsection{Sample of galaxies and phase-space coordinates}
\label{sec:sample}
The sample of galaxies we considered coincides with the sample presented in ~\citet{MartinezGarcia2026}, with the addition of the LMC. It includes 73 dwarf galaxies in total. We briefly outline the sample here, but refer to the mentioned paper for a more detailed description. The sample is limited to dwarfs within 500 kpc of the MW, because galaxies beyond this radius are unlikely to be bound to it. We also restricted the selection to galaxies for which 6D phase-space coordinates, that is, 3D positions (right ascension, declination, and distance) and 3D velocities (PMs and line-of-sight velocity), are available in the literature. When selecting spatial and kinematic data of the MW dwarf neighbours we prioritised works that derived systemic PMs for large numbers of dwarfs in order to have a sample that was as consistent as possible. When not available in such works, galaxies were chosen from individual studies. Our main source of data was \citealt{Battaglia2022}, who performed the largest derivation of systemic PMs of dwarf galaxies to date based on \textit{Gaia} EDR3, and who also provides an exhaustive compilation of positions, distances, and line-of-sight velocities. From this catalogue, we chose a total of 50 galaxies\footnote{We note that unlike for the rest of dwarfs selected from \citealt{Battaglia2022}, for Pisces~II and Tucana~V we adopted the PMs that were derived using the spectroscopic prior, because their derivation without it was flagged as unreliable. Specifically, \citet{Battaglia2022} determined systemic PMs using a maximum-likelihood method based  on the spatial, photometric, and PM information of individual stars. For an additional separate PM determination, they also modelled the line-of-sight velocity distribution of stars with available spectroscopy. For the large majority of their galaxy sample, this addition has a minor effect. For Pisces~II and Tucana~V, however, the PM determination without spectroscopic information is unreliable: for Pisces~II, no stars reach a membership probability above 50\% and the resulting PM posterior is essentially flat, while for Tucana~V, the posterior is strongly lopsided. When the spectroscopic term is included, both issues are resolved and yield a well-constrained posterior in each case. We therefore followed the recommendation of \citet{Battaglia2022} and adopted the spectroscopic-prior PMs for these two galaxies.}
We complemented this sample with Aquarius III, Bootes V, Centaurus I, Eridanus IV, Leo VI, Pegasus III, Pegasus IV, Pictor II, Sagittarius, and the Small Magellanic Cloud (SMC), for which phase-space coordinates were obtained from The Local Volume Database(\citealt{Pace2024})\footnote{Centaurus I, Pictor II, and Pegasus III were already part of \citet{Battaglia2022}, but the line-of-sight velocities of the first two were not available at the time (and were afterwards reported in \citealt{Heiger2024, Pace2025}), while the PM of Pegasus III was flagged as not reliable.}. Finally, we adopted 12 systems from \citet{Cerny2026}, who recently reported phase-space coordinates for 19 ultra-faint compact satellites (UFCSs). The nature of these systems is still debated, and it is unclear whether they are ultra-faint dwarfs (UFDs) or globular clusters. We discarded UFCSs for which the evidence that they are globular clusters was conclusive (labelled as ‘Definite Star Cluster’, ‘Very Likely Star Cluster’,  and ‘Likely Star Cluster’ in Table 5 of \citealt{Cerny2026}), and kept the remainder for our study. 
We note that, we  refer throughout the text to all the considered systems as dwarf galaxies for simplicity, although the nature of some of them is still debated. 
Finally, we included the LMC, for which we adopted $(\alpha, \delta) = (81.28^{\circ}, -69.78^{\circ})$, $\mu_{\alpha} = 1.858$ mas yr$^{-1}$, $\mu_{\delta} = 0.385$ mas yr$^{-1}$ \citep{GaiaCollaboration2021LMC}, $d = 49.6$ kpc \citep{Pietrzynski2019}, and $v_\mathrm{los} = 262.2$ km s$^{-1}$ \citep{vanderMarel2002}. 

Our final sample of 73 systems accounts for nearly 80\% of the 92 confirmed or candidate dwarf galaxies currently known within 500~kpc of the MW \citep{Battaglia2022, Pace2024, Cerny2026}, and the remaining systems were excluded because available 6D phase-space coordinates are lacking. All the kinematic and positional data for the sample we used can be found in Table~\ref{tab:obs_coords} of Appendix~\ref{app:data}, alongside references to the works from which they were obtained.

\subsection{VPOS membership classification}
\label{sec:class}
We followed a  procedure analogous to that described in \citet{Fritz2018} and \citet{Taibi2024} to assess the membership  to the VPOS for the dwarfs in the sample. 
We proceeded as follows for each galaxy. In the first place, we estimated its specific angular momentum. We used a Monte Carlo (MC) scheme to account for all known sources of observational uncertainties, throughout $10^3$ iterations. In each realisation, we sampled the phase-space coordinates of the dwarf from random normal distributions centred on their corresponding nominal values and dispersion equal to their associated uncertainties (when they were asymmetric, we adopted the mean value). Then, we proceeded to transform the sampled phase-space coordinates into galactocentric Cartesian Coordinates using Astropy (\citealt{Astropy2013, Astropy2018}). We used the following parameters for the transformation: $R_{0} = 8.122 \pm 0.021$ kpc (\citealt{GRAVITY2018}), $z_{\odot} = 20.8 \pm 0.3$ pc (\citealt{BennettBovy2019}), $V_{R, \odot} = -12.9 \pm 3.0$ km s$^{-1}$, $V_{\phi, \odot} = 245.6 \pm 1.4$ km s$^{-1}$, and $V_{Z, \odot} = 7.78 \pm 0.09$ km s$^{-1}$ (\citealt{DrimmelPoggio2018}). These solar parameters were likewise randomly sampled in each iteration. With the resulting sampled phase-space coordinates, we  calculated the specific angular momentum, as the cross product of the positions and velocities. We  transformed the direction of the specific angular momentum (i.e. the orbital pole) of each realisation into Galactic coordinates with Astropy. We then estimated the fraction of MC realisations where the orbital poles of a galaxy fell within the area encompassing 10\% of the sky (i.e. a 36.87$^{\circ}$ aperture) around the direction of the normal vector to the VPOS ($l =169.3 ^{\circ}$, $b = -2.8^{\circ}$ \citealt{PawlowskiKroupa2013, Fritz2018}) and its antipode. 
The resulting values of this fraction ($f_{\mathrm{VPOS}}$) quantify the level of alignment of the angular momentum of a dwarf with the VPOS. Dwarfs for which we obtained $f_{\mathrm{VPOS}} \geq 0.95$ were considered on-plane (i.e. they are members of the VPOS), those with $f_{\mathrm{VPOS}} \leq 0.05$ were considered off-plane, and those with $0.05 <f_{\mathrm{VPOS}}< 0.95$ were considered uncertain VPOS members.  

\begin{figure*}
    \centering
    \includegraphics[width=\linewidth]{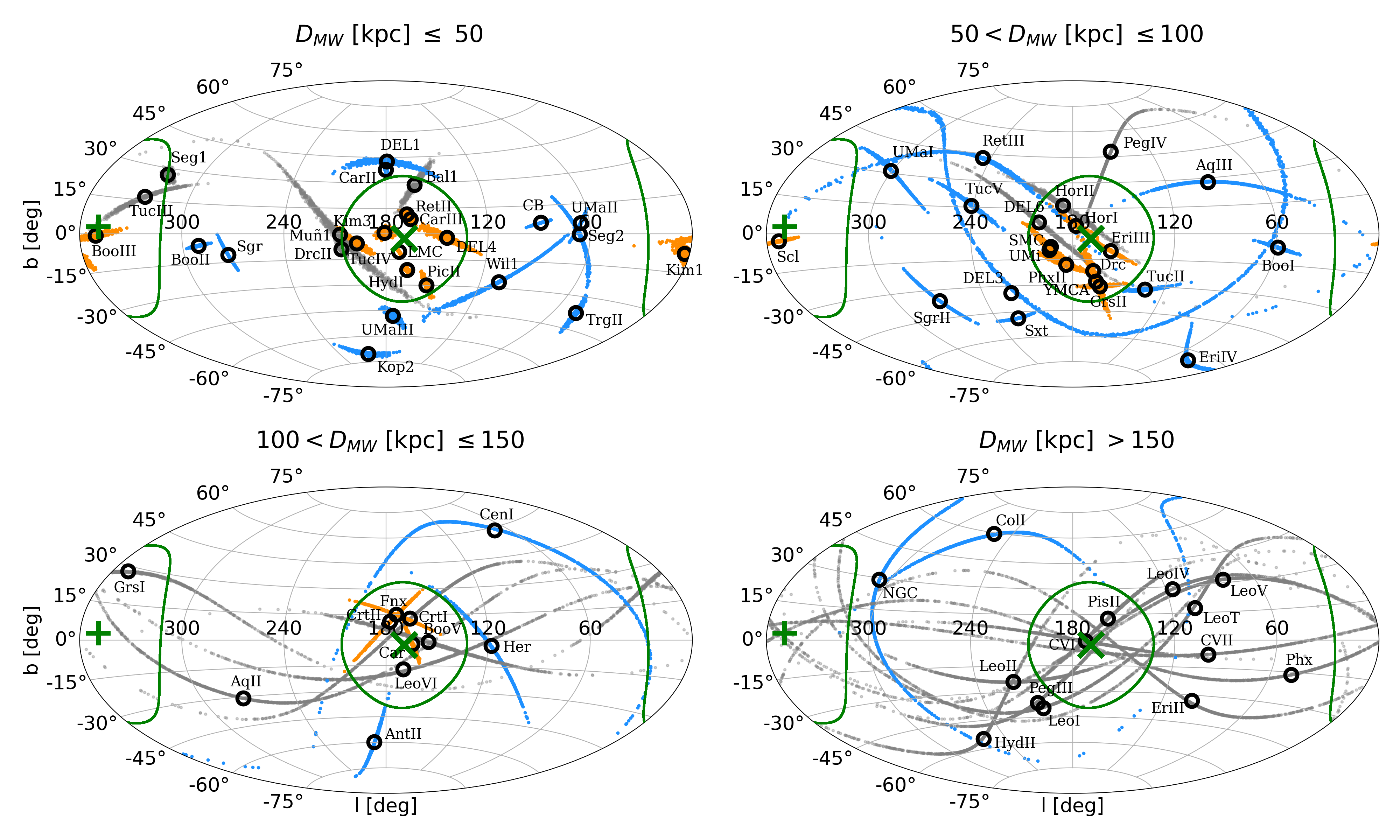}
    \caption{Orbital poles of the 73 MW dwarf galaxies of the sample in Galactic coordinates. The solid coloured points represent the orbital poles obtained for each dwarf from the MC sampling, colour-coded according to their VPOS membership: orange for on-plane dwarfs, blue for off-plane dwarfs, and grey for uncertain members. The empty black circles mark the median orbital pole of each galaxy, labelled with its abbreviated name. The galaxies are distributed in the panels according to their galactocentric distance. The green cross and plus represent the direction of the normal vector to the VPOS and its antipode,  respectively (taken from \citealt{PawlowskiKroupa2013}). The green lines enclose the region encompassing 10\% of the sky around each of them.}
    \label{fig:VPOSmem}
\end{figure*}

\subsection{VPOS structural metrics derivation}
\label{sec:metrics}
We calculated the structural metrics that characterise the VPOS, that is, its planarity, thickness, and orientation, based on the spatial distribution of its member dwarfs (i.e. those with $f_{\mathrm{VPOS}} \geq 0.95$). We conducted this characterisation using a recursive approach to account for the uncertainties in their phase-space coordinates. In each iteration, we sampled the 3D positions of the on-plane dwarfs from random normal distributions and performed a principal component analysis (PCA) by computing the covariance matrix of their resulting spatial distribution. From its eigenvalues we derived the short-to-long axis ratio ($c/a$) and the intermediate-to-long axis ratio ($b/a$). $c/a$ is a measure of the planarity of the VPOS, indicating how thin the distribution of its member galaxies is. Low values imply that the distribution is thin and flattened. Values of $b/a$ close to unity indicate a disc-like  oblate configuration, extended in two dimensions. Conversely, when $b/a$ is small and comparable to $c/a$, the distribution is better described as elongated, or prolate.
The thickness of the VPOS was calculated by projecting the positions of the galaxies along the direction of the shortest principal axis of the galaxy distribution (i.e. the normal vector to the plane, derived from the eigenvectors of the covariance matrix), and calculating the root mean square (RMS) of the projected distances. Although the RMS is the standard method in the literature for estimating the VPOS thickness (e.g. \citealt{Pawlowski2018}), we also measured it based on the median absolute deviation (MAD), which is a more robust alternative that is less sensitive to extreme values. 
Finally, we calculated the direction of the VPOS normal vector. This procedure was repeated in $10^3$ iterations. We then median-averaged the metrics across realisations, and derived their associated uncertainties as the 16th and 84th percentiles of the corresponding distributions.

\section{Results and discussion}
\subsection{VPOS membership}
\label{sec:vposmem}
We assessed the membership to the VPOS of 73 MW dwarf galaxies. 
In Table~\ref{tab:VPOSprob} we present the membership classification of the dwarfs of our sample, alongside the corresponding values of $f_{\mathrm{VPOS}}$. In Figure~\ref{fig:VPOSmem} we show the distribution of their orbital poles in Galactic coordinates.

We found 22 on-plane dwarfs (30\% of the galaxies of the sample), consisting of 12  MW satellites (five classical satellites: Carina, Draco, Fornax, Sculptor, and Ursa Minor; four UFDs: Bootes III, Crater II, Grus II, and Tucana IV; three UFCSs: Delve 4, Kim 1, and Kim 3), nine putative LMC satellites \footnote{We adopted as putative LMC satellites the systems classified as such in \citet{Vasiliev2024} (see the current LMC satellites in Table 3) and \citet{MartinezGarcia2026} (Table 2). In the former, LMC satellite membership is determined using particle matching, Gaussian-mixture modelling, and orbit rewinding, whereas in the latter, LMC satellites are classified as dwarfs with a recursive track of pericentres and apocentres within the tidal radius of the LMC, and also based on their orbital energy and separation to the LMC prior to  crossing the MW virial radius (see full details in the mentioned works). For simplicity, the adopted putative LMC satellites are referred to as LMC satellites throughout.} (the SMC, six UFDs: Carina III, Horologium I, Hydrus I, Phoenix II, Pictor II, and Reticulum II; and two UFCSs: Eridanus III and YMCA-1), and the LMC itself.  
However, we found that the majority of galaxies of our sample are  either off-plane (26 of them, 36\%) or uncertain cases (25 dwarfs or 34\%).
Figure~\ref{fig:VPOSmem} reveals that on-plane galaxies tend to have tightly constrained orbital poles (due to the classification method adopted, see Section~\ref{sec:class}) and are preferentially located at smaller galactocentric distances. Conversely, off-plane and uncertain galaxies typically lie at larger distances and have larger uncertainties in their phase-space coordinates (and in particular, in their PMs) that lead to broader orbital pole distributions. 

The vast majority of the on-plane galaxies are co-rotating in the plane of the VPOS (their orbital poles point in the direction of the VPOS normal vector, see Figure~\ref{fig:VPOSmem}), whereas only three (Bootes III, Kim 1, and Sculptor, 14\% of the on-plane dwarfs) are counter-rotating (their orbital poles lie in the region encompassing  the antipode of the VPOS normal vector).  Similarly high values of co-rotation have previously been reported in observational studies (e.g. \citealt{Fritz2018, Li2021, Taibi2024, MartinezGarcia2025}). However, simulations tend to struggle to reproduce it and generally report lower fractions of co-rotating dwarfs in VPOS analogues (e.g. \citealt{Shao2019, GamezMarin2025, RodriguezCardoso2026}).
The unusually high co-rotation of the on-plane dwarfs might be explained by the presence of the LMC satellite system. The LMC and its associated satellites account for nearly half of the on-plane systems, and the direction of motion of the group matches  the sense of rotation of the majority of MW on-plane satellites (see Figure 2 of \citealt{MartinezGarcia2025}), thus enhancing the co-rotation of the VPOS.  This scenario was supported by \citet{SantosSantos2023}, who found in selected zoom-in simulations that the accretion of LMC-like satellite groups can severely modify the fraction of co-rotating systems in planes of satellites.

The presence of counter-rotating galaxies in the VPOS was also  studied by \citet{Pawlowski2011} through simulations of disc galaxy mergers and fly-bys. They found that counter-rotating debris can be found in the expelled material resulting from such encounters and argued that tidal dwarf galaxies forming from that debris might explain the presence of counter-rotating satellites in the VPOS. The fraction of particles on retrograde orbits in several of their simulations (e.g. fly-by model 6deg180vel, or merger model 7.5deg100vel) is similar to the  fraction of counter-rotating on-plane galaxies found here. However, since \citet{Pawlowski2011} only traced debris particles, neglecting gas dynamics and MW mass growth, and did not investigate satellite formation from the debris, comparisons with our results  do not allow us to draw strong conclusions regarding the formation of the VPOS.

Our results agree  with those reported in \citealt{Li2021},  \citet{Taibi2024}, and \citet{MartinezGarcia2025}, who used the membership classification method we did, but on smaller samples (46, 49, and 58 dwarfs, respectively, all subsumed within our dataset) also based on kinematic data from \textit{Gaia} EDR3. They identified 13, 16, and 15 on-plane satellites, respectively, all of which are recovered as on-plane in our analysis. The only exception is Segue~1, which was only classified as on-plane in \citet{Li2021}, likely because different phase-space coordinates were adopted for this galaxy in the  studies.
In general, the values of $f_{\mathrm{VPOS}}$ obtained in the different studies are also similar to those reported here, and the values derived by \citet{Taibi2024} and \citet{MartinezGarcia2025} agree particularly well  because the underlying phase-space data are very similar (heavily based on \citealt{Battaglia2022}). We note that our analysis reports a significantly larger sample of on-plane dwarfs, having increased  the census of VPOS member galaxies known to date by nearly 50\%. With respect to previous works, we report six new on-plane systems, namely  Delve 4, Kim 1, Kim 3, Pictor II, Eridanus III and YMCA-1. They mostly UFCSs, for which accurate 6D phase-space coordinates were not available until recently. 

Finally, we note that the membership selection method we adopted  relies on a strict boundary classification (Section~\ref{sec:class}), which likely excludes additional candidate members.  A clear example is Carina II, an LMC satellite  whose orbital poles are well constrained and lie immediately outside the boundary for an on-plane classification, showing $f_{\mathrm{VPOS}} = 0$. Its association with the LMC, for which the majority of its satellites belong to the VPOS and share a common motion, further reinforces the idea that Carina II might be a likely on-plane dwarf. Other potential on-plane galaxies are Segue 1 ($f_{\mathrm{VPOS}} = 0.7$) or Draco II ($f_{\mathrm{VPOS}} = 0.26$), whose orbital poles  lie very close to the mentioned boundary. 
We also note that the method we used requires the orbital poles of the dwarfs to be very well constrained. This favours the classification as VPOS members of systems with well-measured phase-space coordinates. When we only consider the median orbital poles of the dwarfs,  they fall within the areas encompassing 10\% of the sky around the VPOS normal vector direction and its antipode for 33 systems (11 more than our on-plane census). Future improvements in positional and kinematic data will tighten the orbital pole distributions, which is expected to increase the VPOS census.

\begin{table}
    \caption{VPOS membership classification}
    \centering
    \fontsize{10}{12}\selectfont
    \setlength{\tabcolsep}{3pt} 
    \begin{tabular}{lcc  lcc}
    \toprule
    \toprule
    Galaxy & Class. & $f_{\mathrm{VPOS}}$ & Galaxy & Class. & $f_{\mathrm{VPOS}}$ \\
    \multicolumn{1}{c}{(1)} & (2) & \multicolumn{1}{c}{(3)} & \multicolumn{1}{c}{(4)} & (5) & \multicolumn{1}{c}{(6)} \\
    \cmidrule(lr){1-3} \cmidrule(lr){4-6} \cmidrule(lr){1-3} \cmidrule(lr){4-6}
    
Antlia II & off & 0.0               & Koposov 2 & off & 0.0             \\    
Aquarius II & ? & 0.14              & Leo I & ? & 0.32                  \\    
Aquarius III & off & 0.0            & Leo II & ? & 0.28                 \\    
Balbinot 1 & ? & 0.77               & Leo IV & ? & 0.24                 \\    
Bootes I & off & 0.0                & Leo T & ? & 0.11                  \\    
Bootes II & off & 0.0               & Leo V & ? & 0.12                  \\    
Bootes III & on & 1.0               & Leo VI & ? & 0.51                 \\    
Bootes V & ? & 0.72                 & LMC & on & 1.0                    \\    
C. Venatici I & ? & 0.80            & Muñoz 1 & ? & 0.44                \\    
C. Venatici II & ? & 0.21           & NGC 6822 & off & 0.0              \\    
Carina & on & 1.0                   & Pegasus III & ? & 0.37            \\    
\textit{Carina II} & off & 0.0      & Pegasus IV & ? & 0.2              \\    
\textit{Carina III} & on & 1.0      & Phoenix & ? & 0.22                \\    
Centaurus I & off & 0.02            & \textit{Phoenix II} & on & 1.0    \\    
Columba I & off & 0.0               & \textit{Pictor II} & on & 0.99    \\    
C. Berenices & off & 0.0            & Pisces II & ? & 0.7               \\    
Crater I & ? & 0.53                 & \textit{Reticulum II} & on & 1.0  \\    
Crater II & on & 1.0                & Reticulum III & off & 0.03        \\    
DELVE 1 & off & 0.0                 & Sagittarius & off & 0.0           \\    
DELVE 3 & off & 0.0                 & Sagittarius II & off & 0.0        \\    
DELVE 4 & on & 0.97                 & Sculptor & on & 1.0               \\    
DELVE 6 & ? & 0.60                  & Segue 1 & ? & 0.7                 \\    
Draco & on & 1.0                    & Segue 2 & off & 0.0               \\    
Draco II & ? & 0.26                 & Sextans & off & 0.0               \\    
Eridanus II & ? & 0.21              & \textit{SMC} & on & 0.99          \\    
\textit{Eridanus III} & on & 1.0    & Triangulum II & off & 0.0         \\    
Eridanus IV & off & 0.0             & Tucana II & off & 0.0             \\    
Fornax & on & 1.0                   & Tucana III & ? & 0.88             \\    
Grus I & ? & 0.62                   & Tucana IV & on & 0.97             \\    
Grus II & on & 1.0                  & Tucana V & off & 0.0              \\    
Hercules & off & 0.0                & Ursa Major I & off & 0.0          \\    
\textit{Horologium I} & on & 1.0    & Ursa Major II & off & 0.0         \\    
\textit{Horologium II} & ? & 0.78   & Ursa Major III & off & 0.0        \\    
Hydra II & ? & 0.06                 & Ursa Minor & on & 1.0             \\    
\textit{Hydrus I} & on & 1.0        & Willman 1 & off & 0.0             \\    
Kim 1 & on & 1.0                    & \textit{YMCA-1} & on & 0.95       \\    
Kim 3 & on & 1.0                    &        &   &                      \\    

    \bottomrule
    \end{tabular}
    \tablefoot{Galaxy sample and their VPOS membership classification. Column 1 displays the galaxies names, Column 2 their classification as on-plane (on), off-plane (off), or uncertain (?), and Column 3 the fraction of MC realizations in which the galaxies orbital poles fall within the area encompassing 10\% of the sky around either the VPOS normal vector direction or its antipode. Columns 4-6 follow the same structure. Those galaxies whose names are written in italics are likely LMC satellites (\citealt{Vasiliev2024, MartinezGarcia2026}).}
    \label{tab:VPOSprob}
\end{table}

\subsection{VPOS key metrics}
Based on the updated census of on-plane dwarfs presented in the previous section, we derived the metrics that characterise the VPOS, namely its axis ratios, thickness, and orientation. The values of these metrics can be found in Table~\ref{tab:metrics}. Given the significant role played by the LMC system in the VPOS census (accounting for nearly half of the on-plane dwarfs, see Section~\ref{sec:vposmem}), we also derived the structural metrics without the LMC and its on-plane satellites (i.e. considering only MW on-plane satellites) in order to assess potential differences in the VPOS geometry, and their implications. 

When considering the whole  on-plane sample, we observed that the axis ratios (high $b/a$ and low $c/a$) confirm the extended planar nature of the VPOS. This is further supported by the remarkably low thickness of the plane, which is 14.5 kpc when measured via the RMS, and even lower (6.7 kpc) when measured via the MAD. The direction of the normal vector to the VPOS in Galactic coordinates shows that the plane is almost perpendicular to the Galactic disc ($b=3.3^{\circ}$). 
 
The structural metrics we derived considering exclusively the MW on-plane satellites vary only mildly with respect to those obtained for the full on-plane sample. The axis ratios reveal minor changes, with the MW-only distribution being slightly less elongated. This is likely a consequence of excluding the ten on-plane dwarfs of the LMC system, which lie approximately 50 kpc from the MW and form a compact group. Their inclusion in the full sample therefore modifies the on-plane distribution towards a more elongated one. This same compactness also affects the thickness of the plane: excluding the LMC dwarfs increases the measured thickness for both the RMS and MAD estimators. The LMC group also affects the orientation of the VPOS. 
The LMC orbit closely matches the plane of the VPOS, which it crosses as it moves through the MW halo. The small misalignment between the two, however, accounts for the $3^{\circ}$ difference in Galactic latitude  between the MW-only and full-sample orientations. In any case, we conclude that the structural parameters of the VPOS remain virtually unchanged regardless of whether on-plane LMC satellites are included.

The VPOS metrics we derived agreed broadly with those reported in previous works (e.g. \citealt{Pawlowski2013, Pawlowski2015, Pawlowski2018, SantosSantos2020b, Pawlowski2021, MartinezGarcia2026}), which usually reported $c/a$ between 0.2-0.3 and an RMS thickness of about 20-30 kpc despite different assumed on-plane samples and methods. Despite 
 the substantial expansion of the on-plane census reported here, the consistency with previous works lends further support to the hypothesis that the VPOS is as a robust and well-defined physical structure, and not a chance alignment arising from small-number statistics.

However, our results disagree with previous studies that reported a spatial extent of approximately 250 kpc for the VPOS (see \citealt{Pawlowski2021} and references therein). This extent results from considering that the VPOS is formed by the 11 classical satellites of the MW\footnote{The 11 classical satellites of the MW are the following: Carina, Draco, Fornax, Leo I, Leo II, Sagittarius, Sculptor, Sextans, Ursa Minor, the LMC, and the SMC (\citealt{Metz2009}).}, among which Leo I is the most distant member, thereby setting the limit  for the extension of the VPOS. In contrast, we did not find Leo I to be an on-plane dwarf (nor Leo II, Sagittarius or Sextans, see Table~\ref{tab:VPOSprob}). Instead, the most distant on-plane dwarf in our census is Fornax, which is located approximately 140 kpc away from the MW (\citealt{Pace2024}), and thus sets the upper limit for the VPOS extension. 

\begin{table}
 \caption{VPOS structural metrics.}
    \centering
    \renewcommand{\arraystretch}{1.4}
    \begin{tabular}{ccc}
    \toprule
    \toprule
    Metric & Full sample & Ex-LMC system \\
        & (22 dwarfs) & (12 dwarfs) \\
    (1) & (2) & (3) \\
    \midrule
    $b/a$ & ${0.601}^{+0.012}_{-0.013}$ & ${0.612}^{+0.013}_{-0.013}$ \\
    $c/a$ & ${0.273}^{+0.006}_{-0.006}$ & ${0.237}^{+0.004}_{-0.005}$ \\
    Thickness (RMS) [kpc] & ${14.5}^{+0.4}_{-0.4}$ & ${15.5}^{+0.3}_{-0.3}$ \\
    Thickness (MAD) [kpc] & ${6.7}^{+0.3}_{-0.3}$ & ${7}^{+1}_{-1}$ \\
    $l$ [deg]   & ${179.4}^{+0.6}_{-0.6}$ & ${179.9}^{+0.5}_{-0.5}$ \\
    $b$ [deg] & ${3.3}^{+0.5}_{-0.6}$ & ${6.2}^{+0.3}_{-0.3}$ \\
    \bottomrule
    
     \end{tabular}
     \tablefoot{Summary of relevant metrics of the VPOS. Columns show, from left to right (1) the derived structural metrics, (2) their values measured using the whole sample of on-plane galaxies and (3) their values measured excluding on-plane galaxies of the LMC system. }
      \label{tab:metrics}
\end{table}

\section{Conclusions}
We presented the largest census of VPOS member galaxies to date. Using a sample of 73 MW dwarf galaxies with 6D phase-space coordinates available in the literature, we evaluated their VPOS membership based on the clustering and alignment of their orbital poles.

We identified 22 dwarf galaxies as members of the VPOS, consisting of 12 MW satellites and 10 dwarfs of the LMC system. Conversely, we found 26 dwarfs that are unambiguously non-members and 25 additional systems with an uncertain association with the VPOS.

Our study increases the census of known VPOS member galaxies by nearly  50\% relative to previous works, reporting six new members: Delve 4, Eridanus III, Kim 1, Kim 3, Pictor II, and YMCA-1. We note that the vast majority of the VPOS member dwarfs co-rotate within the plane, and only three systems are counter-rotating (Bootes III,  Kim 1, and Sculptor). This confirms the extreme coherence of the VPOS.

Based on the updated census of VPOS member galaxies, we re-evaluated its structural metrics. Applying  principal component analysis to the spatial distribution of on-plane dwarfs, we derived the axis ratios, thickness, and orientation of the VPOS. Our results highlight the planarity  ($c/a$ = 0.27) and thin  (15 kpc) configuration of the VPOS, which is oriented almost perpendicular to the disc of the MW, with a small tilt of only $3.3^{\circ}$, in agreement with previous studies.

\begin{acknowledgements}    
The author thanks A. del Pino for discussions on this study.
This work received financial support from the Spanish Ministry of Science and Innovation project PID2024-155572NB-C22 (MICIU/AEI/10.13039/501100011033, FEDER, EU), the RyC-MAX grant 20245MAX008 (CSIC) and the \emph{Severo Ochoa} grant CEX2021-001131-S (MICIU/AEI/10.13039/501100011033).
A. M. Martínez-García thanks A. Almeida for her support and discussions on this work.
\end{acknowledgements}

\bibliographystyle{aa}
\bibliography{biblio.bib}

\clearpage
\onecolumn

\begin{appendix}
\section{Positional and kinematic data}
\label{app:data}
\begingroup
\small
\setlength{\tabcolsep}{6pt}
\renewcommand{\arraystretch}{1.3}
\setlength{\LTleft}{\fill}
\setlength{\LTright}{\fill}
\begin{longtable}{l c c c c c c c}
\caption{Positional and kinematic data for the sample of studied dwarfs} \label{tab:obs_coords} \\
\toprule
Galaxy & $\alpha$ & $\delta$ & $d$ & $\mu_\alpha^*$ & $\mu_\delta$ & $v_\mathrm{los}$ & Ref. \\
 & [deg] & [deg] & [kpc] & [mas yr$^{-1}$] & [mas yr$^{-1}$] & [km s$^{-1}$] & \\
(1) & (2) & (3) & (4) & (5) & (6) & (7) & (8) \\
\midrule
\midrule
\endfirsthead
\multicolumn{8}{c}{{\textbf{\tablename\ \thetable{}} -- Continued}} \\
\toprule
Galaxy & $\alpha$ & $\delta$ & $d$ & $\mu_\alpha^*$ & $\mu_\delta$ & $v_\mathrm{los}$ & Ref. \\
 & [deg] & [deg] & [kpc] & [mas yr$^{-1}$] & [mas yr$^{-1}$] & [km s$^{-1}$] & \\
(1) & (2) & (3) & (4) & (5) & (6) & (7) & (8) \\
\midrule
\midrule
\endhead
\midrule
\multicolumn{8}{r}{\textit{Continued on next page...}} \\
\endfoot
\midrule
\multicolumn{8}{c}{%
  \parbox{0.85\textwidth}{\vspace{4pt}\small \textbf{Note.} Positional and kinematic data for the galaxies of the sample.  Columns show, from left to right: (1) galaxy name; (2) Right Ascension ($\alpha$); (3) Declination ($\delta$); (4) heliocentric distance ($d$); (5) PM in Right Ascension direction ($\mu_\alpha^*$); (6) PM in Declination direction ($\mu_\delta$); (7) line-of-sight velocity ($v_\mathrm{los}$); and (8)  reference from where data was selected. \textbf{References.} (1) \citet{Battaglia2022}; (2) \citet{Pace2025}; (3) \citet{Cerny2026}; (4) \citet{GaiaCollaboration2021LMC} ; (5) \citet{Pietrzynski2019}; and (6) \citet{vanderMarel2002} (see references therein).}
} \\
\endlastfoot
Antlia II & $143.8867$ & $-36.7673$ & $132^{+7}_{-7}$ & $-0.100^{+0.007}_{-0.007}$ & $0.087^{+0.008}_{-0.007}$ & $290.7^{+0.5}_{-0.5}$ & (1) \\
Aquarius II & $338.4813$ & $-9.3274$ & $108^{+4}_{-3}$ & $-0.0^{+0.2}_{-0.2}$ & $-0.5^{+0.2}_{-0.2}$ & $-71^{+3}_{-3}$ & (1) \\
Aquarius III & $357.2180$ & $-3.4890$ & $86^{+4}_{-4}$ & $1.0^{+0.3}_{-0.3}$ & $-0.1^{+0.2}_{-0.2}$ & $-13.1^{+1.0}_{-0.9}$ & (2) \\
Balbinot 1 & $332.6800$ & $14.9400$ & $32^{+1}_{-2}$ & $1.5^{+0.1}_{-0.1}$ & $-1.3^{+0.1}_{-0.1}$ & $-176^{+1}_{-2}$ & (3) \\
Bootes I & $210.0225$ & $14.5006$ & $66^{+2}_{-2}$ & $-0.39^{+0.02}_{-0.02}$ & $-1.06^{+0.01}_{-0.01}$ & $101.8^{+0.7}_{-0.7}$ & (1) \\
Bootes II & $209.5179$ & $12.8525$ & $42^{+1}_{-1}$ & $-2.46^{+0.07}_{-0.07}$ & $-0.45^{+0.05}_{-0.06}$ & $-117^{+5}_{-5}$ & (1) \\
Bootes III & $209.3000$ & $26.8000$ & $47^{+2}_{-2}$ & $-1.16^{+0.02}_{-0.02}$ & $-0.88^{+0.01}_{-0.01}$ & $198^{+4}_{-4}$ & (1) \\
Bootes V & $213.9090$ & $32.9140$ & $102^{+7}_{-7}$ & $-0.22^{+0.05}_{-0.05}$ & $-0.28^{+0.07}_{-0.07}$ & $5^{+13}_{-13}$ & (2) \\
Canes Venatici I & $202.0091$ & $33.5521$ & $211^{+5}_{-5}$ & $-0.10^{+0.03}_{-0.03}$ & $-0.12^{+0.02}_{-0.02}$ & $30.9^{+0.6}_{-0.6}$ & (1) \\
Canes Venatici II & $194.2927$ & $34.3226$ & $160^{+4}_{-4}$ & $-0.1^{+0.1}_{-0.1}$ & $-0.29^{+0.08}_{-0.08}$ & $-129^{+1}_{-1}$ & (1) \\
Carina & $100.4065$ & $-50.9593$ & $106^{+6}_{-6}$ & $0.530^{+0.006}_{-0.006}$ & $0.127^{+0.006}_{-0.006}$ & $222.9^{+0.1}_{-0.1}$ & (1) \\
Carina II & $114.1067$ & $-57.9991$ & $36.1^{+0.8}_{-0.8}$ & $1.88^{+0.02}_{-0.01}$ & $0.13^{+0.01}_{-0.02}$ & $477^{+1}_{-1}$ & (1) \\
Carina III & $114.6298$ & $-57.8997$ & $28^{+1}_{-1}$ & $3.10^{+0.03}_{-0.04}$ & $1.39^{+0.04}_{-0.04}$ & $285^{+3}_{-3}$ & (1) \\
Centaurus I & $189.5908$ & $-40.9043$ & $118^{+4}_{-4}$ & $-0.14^{+0.05}_{-0.05}$ & $-0.19^{+0.04}_{-0.04}$ & $44.8^{+0.8}_{-0.8}$ & (2) \\
Columba I & $82.8570$ & $-28.0425$ & $183^{+9}_{-9}$ & $0.17^{+0.07}_{-0.07}$ & $-0.41^{+0.08}_{-0.08}$ & $154^{+5}_{-5}$ & (1) \\
Coma Berenices & $186.7458$ & $23.9069$ & $42^{+2}_{-2}$ & $0.43^{+0.02}_{-0.02}$ & $-1.72^{+0.02}_{-0.02}$ & $98.1^{+0.9}_{-0.9}$ & (1) \\
Crater I & $174.0660$ & $-10.8778$ & $145^{+3}_{-3}$ & $-0.0^{+0.1}_{-0.1}$ & $-0.12^{+0.10}_{-0.10}$ & $149^{+1}_{-1}$ & (1) \\
Crater II & $177.3280$ & $-18.4180$ & $117^{+4}_{-4}$ & $-0.07^{+0.02}_{-0.02}$ & $-0.11^{+0.01}_{-0.01}$ & $87.5^{+0.4}_{-0.4}$ & (1) \\
DELVE 1 & $247.7250$ & $-0.9720$ & $22^{+2}_{-2}$ & $0.02^{+0.07}_{-0.07}$ & $-1.54^{+0.05}_{-0.05}$ & $-403^{+1}_{-1}$ & (3) \\
DELVE 3 & $290.3960$ & $-60.7840$ & $56^{+2}_{-6}$ & $-0.3^{+0.3}_{-0.3}$ & $-0.8^{+0.3}_{-0.3}$ & $-85^{+3}_{-3}$ & (3) \\
DELVE 4 & $230.7750$ & $27.3950$ & $45^{+4}_{-4}$ & $0.42^{+0.08}_{-0.08}$ & $-0.8^{+0.1}_{-0.1}$ & $-168^{+1}_{-1}$ & (3) \\
DELVE 6 & $33.0700$ & $-66.0560$ & $80^{+4}_{-6}$ & $0.9^{+0.4}_{-0.4}$ & $-1.3^{+0.4}_{-0.4}$ & $171^{+7}_{-6}$ & (3) \\
Draco & $260.0684$ & $57.9185$ & $81^{+3}_{-3}$ & $0.044^{+0.005}_{-0.005}$ & $-0.187^{+0.006}_{-0.006}$ & $-292.3^{+0.4}_{-0.4}$ & (1) \\
Draco II & $238.1740$ & $64.5790$ & $21.6^{+0.5}_{-0.5}$ & $1.12^{+0.09}_{-0.09}$ & $0.91^{+0.10}_{-0.10}$ & $-343^{+1}_{-1}$ & (1) \\
Eridanus II & $56.0838$ & $-43.5338$ & $363^{+17}_{-16}$ & $0.2^{+0.1}_{-0.1}$ & $0.0^{+0.1}_{-0.1}$ & $76^{+3}_{-3}$ & (1) \\
Eridanus III & $35.6950$ & $-52.2840$ & $91^{+4}_{-4}$ & $1.36^{+0.08}_{-0.07}$ & $-0.64^{+0.08}_{-0.08}$ & $54^{+3}_{-2}$ & (3) \\
Eridanus IV & $76.4246$ & $-9.5189$ & $70^{+4}_{-3}$ & $0.25^{+0.06}_{-0.06}$ & $-0.10^{+0.05}_{-0.05}$ & $-32^{+1}_{-1}$ & (2) \\
Fornax & $39.9667$ & $-34.5136$ & $139^{+3}_{-3}$ & $0.381^{+0.001}_{-0.001}$ & $-0.358^{+0.002}_{-0.002}$ & $54.1^{+0.5}_{-0.5}$ & (1) \\
Grus I & $344.1660$ & $-50.1680$ & $126^{+6}_{-6}$ & $0.07^{+0.05}_{-0.05}$ & $-0.27^{+0.08}_{-0.07}$ & $-141^{+2}_{-2}$ & (1) \\
Grus II & $331.0250$ & $-46.4420$ & $55^{+2}_{-2}$ & $0.39^{+0.03}_{-0.04}$ & $-1.51^{+0.04}_{-0.04}$ & $-110.0^{+0.5}_{-0.5}$ & (1) \\
Hercules & $247.7722$ & $12.7852$ & $137^{+11}_{-10}$ & $-0.04^{+0.05}_{-0.05}$ & $-0.34^{+0.04}_{-0.04}$ & $45^{+1}_{-1}$ & (1) \\
Horologium I & $43.8813$ & $-54.1160$ & $79^{+8}_{-7}$ & $0.85^{+0.04}_{-0.03}$ & $-0.61^{+0.04}_{-0.04}$ & $113^{+3}_{-3}$ & (1) \\
Horologium II & $49.1077$ & $-50.0486$ & $78^{+8}_{-7}$ & $1.0^{+0.2}_{-0.2}$ & $-0.8^{+0.2}_{-0.2}$ & $169^{+13}_{-13}$ & (1) \\
Hydra II & $185.4251$ & $-31.9860$ & $151^{+9}_{-8}$ & $-0.4^{+0.1}_{-0.1}$ & $-0.0^{+0.1}_{-0.1}$ & $303^{+1}_{-1}$ & (1) \\
Hydrus I & $37.3892$ & $-79.3089$ & $27.5^{+0.5}_{-0.5}$ & $3.79^{+0.01}_{-0.01}$ & $-1.50^{+0.01}_{-0.01}$ & $80.4^{+0.6}_{-0.6}$ & (1) \\
Kim 1 & $332.9210$ & $7.0270$ & $20^{+1}_{-1}$ & $-1.1^{+0.1}_{-0.1}$ & $-0.09^{+0.09}_{-0.09}$ & $-261^{+1}_{-1}$ & (3) \\
Kim 3 & $200.6880$ & $-30.6000$ & $15.1^{+1.0}_{-0.3}$ & $-0.9^{+0.2}_{-0.2}$ & $3.3^{+0.1}_{-0.1}$ & $163.2^{+0.7}_{-0.8}$ & (3) \\
Koposov 2 & $119.5720$ & $26.2570$ & $24^{+2}_{-2}$ & $-0.6^{+0.2}_{-0.2}$ & $0.1^{+0.1}_{-0.1}$ & $110^{+2}_{-1}$ & (3) \\
Leo I & $152.1146$ & $12.3059$ & $269^{+13}_{-12}$ & $-0.06^{+0.01}_{-0.01}$ & $-0.117^{+0.009}_{-0.010}$ & $282.9^{+0.5}_{-0.5}$ & (1) \\
Leo II & $168.3627$ & $22.1529$ & $217^{+11}_{-11}$ & $-0.11^{+0.03}_{-0.03}$ & $-0.14^{+0.03}_{-0.03}$ & $78.3^{+0.6}_{-0.6}$ & (1) \\
Leo IV & $173.2405$ & $-0.5453$ & $154^{+5}_{-5}$ & $-0.0^{+0.1}_{-0.1}$ & $-0.3^{+0.1}_{-0.1}$ & $131^{+1}_{-1}$ & (1) \\
Leo T & $143.7292$ & $17.0482$ & $409^{+29}_{-27}$ & $0.2^{+0.4}_{-0.4}$ & $-0.1^{+0.2}_{-0.2}$ & $38^{+2}_{-2}$ & (1) \\
Leo V & $172.7857$ & $2.2194$ & $178^{+7}_{-6}$ & $0.1^{+0.2}_{-0.2}$ & $-0.4^{+0.2}_{-0.2}$ & $173.0^{+1.0}_{-0.8}$ & (1) \\
Leo VI & $171.0770$ & $24.8740$ & $111^{+4}_{-6}$ & $-0.1^{+0.2}_{-0.2}$ & $-0.2^{+0.2}_{-0.2}$ & $170^{+2}_{-1}$ & (2) \\
LMC & $81.2800$ & $-69.7800$ & $49.6^{+0.6}_{-0.6}$ & $1.86^{+0.02}_{-0.02}$ & $0.39^{+0.02}_{-0.02}$ & $262^{+3}_{-3}$ & (4), (5), (6) \\
Muñoz 1 & $225.4500$ & $66.9690$ & $45^{+5}_{-5}$ & $-0.0^{+0.6}_{-0.6}$ & $0.8^{+0.6}_{-0.6}$ & $-139^{+5}_{-4}$ & (3) \\
NGC 6822 & $296.2358$ & $-14.7892$ & $470^{+38}_{-35}$ & $-0.058^{+0.009}_{-0.009}$ & $-0.069^{+0.007}_{-0.008}$ & $-55^{+2}_{-2}$ & (1) \\
Pegasus III & $336.1074$ & $5.4150$ & $215^{+12}_{-12}$ & $-0.0^{+0.2}_{-0.2}$ & $-0.6^{+0.2}_{-0.2}$ & $-260^{+1}_{-1}$ & (2) \\
Pegasus IV & $328.5390$ & $26.6200$ & $90^{+1}_{-1}$ & $0.33^{+0.07}_{-0.07}$ & $-0.21^{+0.08}_{-0.08}$ & $-271^{+1}_{-1}$ & (2) \\
Phoenix & $27.7763$ & $-44.4447$ & $409^{+23}_{-22}$ & $0.08^{+0.03}_{-0.03}$ & $-0.06^{+0.04}_{-0.04}$ & $-21^{+1}_{-1}$ & (1) \\
Phoenix II & $354.9928$ & $-54.4050$ & $83^{+8}_{-7}$ & $0.50^{+0.05}_{-0.05}$ & $-1.20^{+0.06}_{-0.06}$ & $32^{+4}_{-4}$ & (1) \\
Pictor II & $101.1750$ & $-59.9010$ & $45^{+2}_{-2}$ & $1.16^{+0.05}_{-0.05}$ & $1.18^{+0.05}_{-0.05}$ & $327^{+1}_{-1}$ & (2) \\
Pisces II & $344.6345$ & $5.9526$ & $183^{+15}_{-14}$ & $0.7^{+0.4}_{-0.4}$ & $-0.6^{+0.3}_{-0.2}$ & $-227^{+3}_{-3}$ & (1) \\
Reticulum II & $53.9493$ & $-54.0466$ & $32^{+1}_{-1}$ & $2.37^{+0.02}_{-0.02}$ & $-1.35^{+0.02}_{-0.02}$ & $64.7^{+1.3}_{-0.8}$ & (1) \\
Reticulum III & $56.3600$ & $-60.4500$ & $92^{+14}_{-12}$ & $0.3^{+0.1}_{-0.1}$ & $-0.6^{+0.2}_{-0.3}$ & $274^{+8}_{-7}$ & (1) \\
Sagittarius & $284.0951$ & $-30.5499$ & $26^{+2}_{-2}$ & $-2.679^{+0.001}_{-0.001}$ & $-1.394^{+0.001}_{-0.001}$ & $140^{+2}_{-2}$ & (2) \\
Sagittarius II & $298.1647$ & $-22.0651$ & $69^{+7}_{-6}$ & $-0.77^{+0.04}_{-0.04}$ & $-0.91^{+0.02}_{-0.02}$ & $-177.2^{+0.5}_{-0.6}$ & (1) \\
Sculptor & $15.0183$ & $-33.7186$ & $84^{+2}_{-2}$ & $0.099^{+0.002}_{-0.002}$ & $-0.159^{+0.002}_{-0.002}$ & $110.6^{+0.5}_{-0.5}$ & (1) \\
Segue 1 & $151.7504$ & $16.0756$ & $23^{+2}_{-2}$ & $-2.06^{+0.05}_{-0.05}$ & $-3.42^{+0.05}_{-0.04}$ & $208.5^{+0.9}_{-0.9}$ & (1) \\
Segue 2 & $34.8226$ & $20.1624$ & $34^{+3}_{-3}$ & $1.43^{+0.06}_{-0.05}$ & $-0.31^{+0.04}_{-0.05}$ & $-40.2^{+0.9}_{-0.9}$ & (1) \\
Sextans & $153.2680$ & $-1.6180$ & $84.7^{+0.4}_{-0.4}$ & $-0.401^{+0.008}_{-0.008}$ & $0.022^{+0.009}_{-0.009}$ & $226.0^{+0.6}_{-0.6}$ & (1) \\
SMC & $16.2500$ & $-72.4200$ & $63^{+3}_{-3}$ & $0.8^{+0.1}_{-0.1}$ & $-1.21^{+0.03}_{-0.03}$ & $145.6^{+0.6}_{-0.6}$ & (2) \\
Triangulum II & $33.3252$ & $36.1702$ & $28^{+1}_{-1}$ & $0.58^{+0.05}_{-0.05}$ & $0.08^{+0.06}_{-0.06}$ & $-382^{+1}_{-1}$ & (1) \\
Tucana II & $342.9796$ & $-58.5689$ & $58^{+6}_{-5}$ & $0.91^{+0.02}_{-0.02}$ & $-1.27^{+0.03}_{-0.03}$ & $-129^{+4}_{-4}$ & (1) \\
Tucana III & $359.1075$ & $-59.5833$ & $23^{+1}_{-1}$ & $-0.08^{+0.02}_{-0.02}$ & $-1.63^{+0.02}_{-0.02}$ & $-101.2^{+0.5}_{-0.5}$ & (1) \\
Tucana IV & $0.7170$ & $-60.8300$ & $47^{+4}_{-4}$ & $0.56^{+0.05}_{-0.05}$ & $-1.69^{+0.05}_{-0.05}$ & $16^{+2}_{-2}$ & (1) \\
Tucana V & $354.3470$ & $-63.2660$ & $55^{+3}_{-8}$ & $-0.14^{+0.04}_{-0.05}$ & $-1.18^{+0.05}_{-0.06}$ & $-36^{+3}_{-2}$ & (1) \\
Ursa Major I & $158.7706$ & $51.9479$ & $97^{+6}_{-6}$ & $-0.40^{+0.03}_{-0.03}$ & $-0.63^{+0.04}_{-0.04}$ & $-55^{+1}_{-1}$ & (1) \\
Ursa Major II & $132.8726$ & $63.1335$ & $35^{+3}_{-2}$ & $1.73^{+0.02}_{-0.02}$ & $-1.90^{+0.02}_{-0.02}$ & $-117^{+2}_{-2}$ & (1) \\
Ursa Major III & $174.7080$ & $31.0780$ & $10^{+1}_{-1}$ & $-0.76^{+0.09}_{-0.08}$ & $1.2^{+0.1}_{-0.1}$ & $89.0^{+0.6}_{-0.6}$ & (3) \\
Ursa Minor & $227.2420$ & $67.2221$ & $76^{+4}_{-4}$ & $-0.124^{+0.004}_{-0.004}$ & $0.071^{+0.005}_{-0.005}$ & $-246.9^{+0.4}_{-0.4}$ & (1) \\
Willman 1 & $162.3436$ & $51.0501$ & $38^{+8}_{-6}$ & $0.28^{+0.06}_{-0.07}$ & $-1.11^{+0.08}_{-0.08}$ & $-14^{+1}_{-1}$ & (1) \\
YMCA-1 & $110.8380$ & $-64.8320$ & $55^{+4}_{-4}$ & $0.9^{+0.2}_{-0.2}$ & $1.2^{+0.2}_{-0.2}$ & $362^{+2}_{-2}$ & (3) \\
\bottomrule
\end{longtable}
\endgroup
\end{appendix}

\twocolumn
\end{document}